\documentclass[a4paper,notitlepage,twocolumn,footnoteinbib,longbibliography,superscriptaddress,10pt]{revtex4-1}

\usepackage{amsmath,amssymb,amsthm,mathtools,color,listings,graphicx}
\usepackage[colorlinks=true,linkcolor=blue,citecolor=blue]{hyperref}

\DeclareMathAlphabet{\mathbbb}{U}{bbold}{m}{n}

\begin{document}

\title{Hyperscaling in the coherent hyperspin machine}
\author{Marcello Calvanese Strinati}
\email{marcello.calvanesestrinati@gmail.com}
\affiliation{Centro Ricerche Enrico Fermi (CREF), Via Panisperna 89a, 00184 Rome, Italy}
\author{Claudio Conti}
\affiliation{Physics Department, Sapienza University of Rome, 00185 Rome, Italy}
\date{\today}

\begin{abstract}
Classical or quantum physical systems can simulate the Ising Hamiltonian for large-scale optimization and machine learning. However, devices such as quantum annealers and coherent Ising machines suffer an exponential drop in the probability of success in finite-size scaling. We show that by exploiting high dimensional embedding of the Ising Hamiltonian and subsequent dimensional annealing, the drop is counteracted by an exponential improvement in the performance. Our analysis relies on extensive statistics of the convergence dynamics by high-performance computing. We propose a realistic experimental implementation of the new annealing device by off-the-shelf coherent Ising machine technology. The hyperscaling heuristics can also be applied to other quantum or classical Ising machines by engineering nonlinear gain, loss, and non-local couplings.
\end{abstract}

\maketitle

Complex optimization problems permeate science and society and are crucial to machine learning~\cite{date2021qubo}, traffic and portfolio optimization~\cite{01605682.2020.1718019}, markets and finance~\cite{khanna2017}, life science~\cite{zhanglifecience}, bioinformatics~\cite{naturecombinatorial}, protein folding~\cite{winfree2002protein}, and epidemic spreading~\cite{zhou2020minimization}. Finding efficient ways to tackle such problems means determining the optimal configuration of a massive amount of degrees of freedom interacting in a highly nontrivial way. Building a generation of computing machines to tackle large-scale optimization is one of the most significant challenges of modern science.

A viable route is offered by the possibility of mapping optimization problems onto a classical Ising Hamiltonian~\cite{10.3389/fphy.2014.00005}. Solving the hard optimization problem then translates into finding the ground state (GS) of the corresponding Ising system. Quantum annealers (QAs) as D-WAVE achieve this mapping by superconducting technology~\cite{Harris2018, King2022}. The last QAs generation reaches the scale of $5\,000$ qubits~\cite{King2023}. Coherent Ising machines (CIMs)~\cite{Marandi2016} exploit optical pulses in a network of degenerate parametric oscillators (POs) simulating up to $100\,000$ spins with tunable couplings~\cite{Honjo2021}.

Both these groundbreaking technologies suffer poor scaling with the number of spins $N$. The success probability drops more than exponentially with $N$ for QAs, while a better performance has been reported for the CIMs~\cite{1805.05217}. When $N$ grows, the number of local minima grows exponentially for a spin glass. Correspondingly, the probability of being stuck in a sub-optimal solution increases. This mechanism lies at the origin of the poor finite-size scaling of Ising machines and optimization algorithms.

During a run in an annealing device, one carefully chooses specific hyperparameters as the pump power $h$ of CIMs. A large number of local minima narrows the range of $h$ where the ground state is found~\cite{PhysRevLett.126.143901}. The difference between the final energy $E$ after minimization and the target ground-state energy $E_{\rm GS}$ signals the absence of convergence. A strategy to overcome these detrimental effects is increasing dimensionality, for example, by replacing the binary Ising model with continuous spin systems. Considering one local minimum in the phase space, additional dimensions turn the minimum into a saddle and open escape directions. However, this also affects the global minimum, whose energy lowers when the number of dimensions increases. By increasing the dimensionality, one escapes local minima but loses the opportunity to identify the target (Ising) ground state.

We introduced the idea of hyperspins and dimensional annealing to exploit large dimensional spaces~\cite{strinati2022hyperspinmachine}. One starts from the binary Ising model encoding a graph defined by a coupling matrix $\mathbf{J}_N$. The binary spins are replaced by unitary vectors in a $D$-dimensional space (the \emph{hyperspins}) on a hypergraph with the same $\mathbf{J}_N$. The hyperspin time-evolution leads to a steady-state energy $E<E_{\rm GS}$. After reaching the steady state, the topology of the hypergraph morphs adiabatically to the original binary Ising model, a heuristic that we name \emph{dimensional annealing}. Notably, the approach radically widens the parameter range of successful convergence to the Ising GS for dense and computationally hard graphs~\cite{strinati2022hyperspinmachine}.

However, the way to physically realize hyperspins and dimensional annealing has yet to be considered. Authors reported on proposals relying on nonlinear POs~\cite{PhysRevApplied.16.054022, yonezu202310ghzclock} and the first observation of hyperspin dynamics~\cite{Ben-Ami:23}. Here we show that a simple modification to the feedback mechanism in the conventional CIM allows the implementation of the hyperspin heuristics.

In this paper, we validate the resulting coherent hyperspin machine (CHYM) by ab-initio, first-principle parallel large-scale simulations. We also compare with the XY machine~\cite{Berloff2017,Gershenzon2020}. We show an exponential increase in the probability of success, with one order of magnitude improvement in the convergence range and the accuracy of the ground-state energy.

\begin{figure}[t]
\centering
\includegraphics[width=7.5cm]{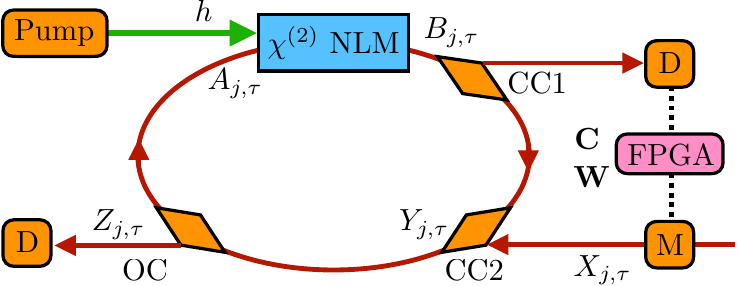}
\caption{Scheme of the CHYM. A $\chi^{(2)}$ NLM pumped by a field at amplitude $h$ and frequency $2\omega_0$ (green) provides amplification of the PO fields $A_{j,\tau}$ at frequency $\omega_0$ (red). Two couplers (CC1 and CC2) connect the optical cavity to the electronic feedback composed by detector (D), FPGA, and modulator (M), which digitally implements both the linear ($\mathbf{C}$) and nonlinear ($\mathbf{W}$) coupling. A output coupler (OC) sends the signal to the measurement device (D). The fields $A_{j,\tau}$, $B_{j,\tau}$, $X_{j,\tau}$, $Y_{j,\tau}$, and $Z_{j,\tau}$ after the respective blocks are shown.}
\label{fig:hyperrcimpaperfigure1}
\end{figure}

We start by introducing the CHYM following the hardware of the CIM~\cite{1805.05217,bohm2019,PhysRevApplied.19.L031001} with a different field programmable gate array (FPGA) including linear and nonlinear coupling between POs. As shown in Fig.~\ref{fig:hyperrcimpaperfigure1}, a $\chi^{(2)}$ nonlinear medium (NLM), pumped by a field at amplitude $h$ and frequency $2\omega_0$, amplifies the PO degenerate fields $A_{j,\tau}$ at frequency $\omega_0$. The phase-dependent amplification enforces a phase $0$ or $\pi$ with respect to the pump, such that $A_{j,\tau}$ is real valued~\cite{PhysRevApplied.16.054022}. Here $j=1,\ldots,D\times N$ labels the POs where $N$ and $D$ are the number of hyperspins and the hyperspin dimension, respectively, and $\tau=1,\ldots,\tau_{\rm max}$ counts the cavity round trip. The NLM takes $A_{j,\tau}$ as input, and yields $B_{j,\tau}=K^{(j)}_\tau A_{j,\tau}$ as output, where $K^{(j)}_\tau>0$ describes parametric amplification (see supplementary material for details).

The coupling device with two identical couplers (CC1 and CC2), a FPGA, a detector (D), and a modulator (M), realizes the organization of the $D\times N$ PO fields as $N$, $D$-dimensional, hyperspins and their mutual coupling. The coupler CC1 extracts a fraction $\sqrt{b}<1$ of the amplified fields $B_{j,\tau}$. Then, the FPGA computes the feedback fields according to
\begin{equation}
 X_{j,\tau}\!=\!\sqrt{b}\left(\sum_{l=1}^{DN}C_{jl}B_{l,\tau}-bB_{j,\tau}\sum_{r=1}^{DN}W_{jl}B_{r,\tau}^2\right) \,\, .
\label{eq:hyperspinmap1}
\end{equation}
In Eq.~\eqref{eq:hyperspinmap1}, the matrix $\mathbf{W}=\beta\mathbbb{1}_N\otimes\mathcal{I}_D$ describes an effective nonlinear loss for the POs, where $\mathbbb{1}_N$ is the $N\times N$ identity matrix, $\mathcal{I}_D$ is the $D\times D$ matrix with all entries equal to one, and $\beta$ is a rescaling factor. This nonlinear loss in the FPGA induces the hyperspins. The linear coupling between hyperspins is described by a matrix $\mathbf{C}=\mathbf{J}_N\otimes\mathbbb{1}_D$, where $\mathbf{J}_N$ is the $N\times N$ symmetric adjacency matrix defining the hyperspin graph~\cite{strinati2022hyperspinmachine}.

The fields $X_{j,\tau}$ in Eq.~\eqref{eq:hyperspinmap1} modulate (M in Fig.~\ref{fig:hyperrcimpaperfigure1}) the optical field injected back into the cavity. After CC2, the coupled field is ($a+b=1$)
\begin{equation}
Y_{j,\tau}\!=\!aB_{j,\tau}+\sqrt{b}X_{j,\tau}\!=\!\sum_{l=1}^{DN}Q_{jl}B_{l,\tau}-b^2B_{j,\tau}\!\sum_{r=1}^{DN}W_{jr}B_{r,\tau}^2 \,\, ,
\label{eq:hyperspinmap2}
\end{equation}
where $\mathbf{Q}=a\mathbbb{1}+b\mathbf{C}$. Then, $Z_{j,\tau}=dY_{j,\tau}$ with $d<1$ is extracted by the output coupler (OC), so $A_{j,\tau+1}=(1-d)Y_{j,\tau}$ is the field at the subsequent round trip. This recurrence relation allows to describe the CHYM dynamics by a discrete-time nonlinear map~\cite{strogatz2007nonlinear} (see~\cite{Calvanese_Strinati_2020,PhysRevLett.126.143901,PhysRevApplied.16.054022} for previous studies on Ising machines).

\begin{figure}[t]
\centering
\includegraphics[width=8.6cm]{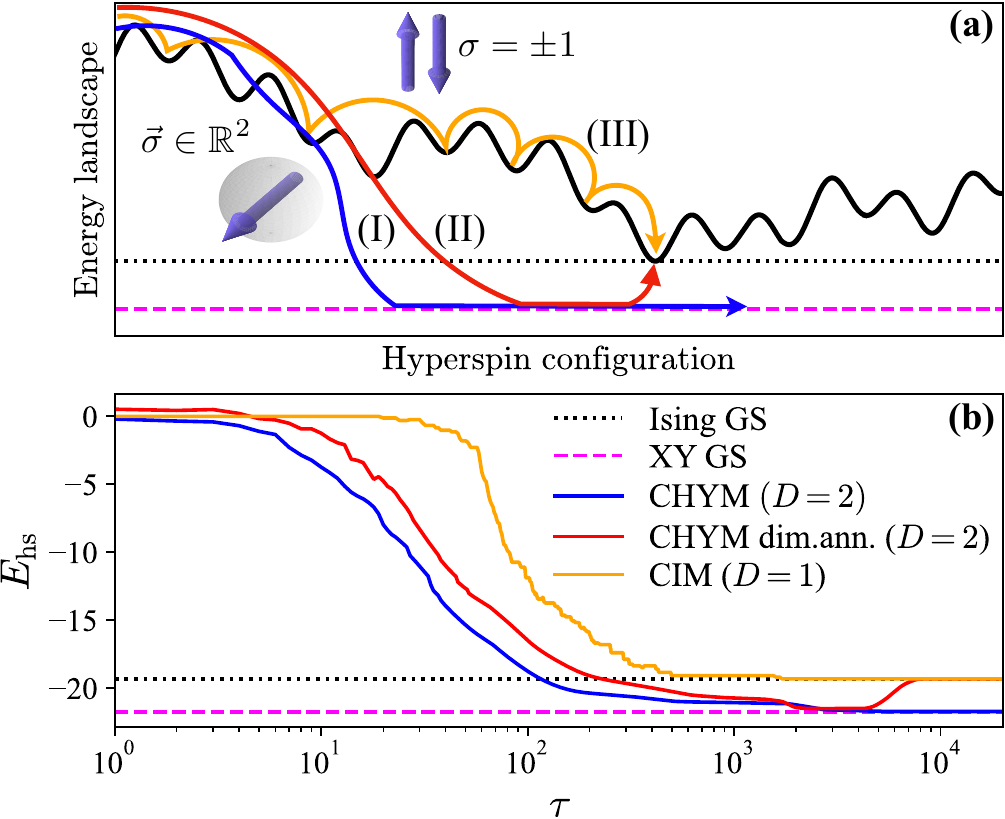}
\caption{\textbf{(a)} Sketch of energy dynamics on the Ising energy landscape showing the principle of operation of the CHYM working as (I) XY machine ($D=2$, blue line), (II) XY machine with dimensional annealing (from $D=2$ to $D=1$, red line), and (III) Ising machine ($D=1$, orange line). Full black wavy line depicts the Ising energy landscape for the CIM. Horizontal dotted black and dashed magenta line mark the minimum (GS) energies for the Ising and XY model, respectively. The CIM approaches the Ising GS from larger energy values, visiting local minima by flipping the Ising spins $\sigma=\pm1$ (arrows) during the evolution that can trap the dynamics. On the contrary, the CHYM simulates continuous spins $\vec{\sigma}\in\mathbb{R}^2$ (arrow in the sphere) and is insensitive to the Ising local minima. The energy tunnels straight towards the XY GS, which is below the Ising GS. From the XY GS, the Ising GS energy is approached from below by the dimensional annealing. \textbf{(b)} Hyperspin energy $E_{\rm hs}$ in Eq.~\eqref{eq:hyperspinmap3} during the round trips $\tau$ from a CHYM simulation, confirming the picture in panel \textbf{(a)}.}
\label{fig:hyperrcimpaperfigure2}
\end{figure}

Figure~\ref{fig:hyperrcimpaperfigure2} shows the CHYM working principle, with the time evolution of the hyperspin energy (see supplementary material)
\begin{equation}
E_{\rm hs}=-\sum_{p,q=1}^{N}J_{pq}\,\vec{\sigma}_{p,\tau}\cdot\vec{\sigma}_{q,\tau} \,\, ,
\label{eq:hyperspinmap3}
\end{equation}
for (I) CHYM (simulating the planar XY model~\cite{PhysRevLett.20.589} with $D=2$, blue line), (II) CHYM with dimensional annealing (interpolating between $D=2$ and $D=1$, red line), and (III) CIM ($D=1$, orange line). During the time evolution, the system explores different spin configurations behaving as gradient descent in the energy landscape of the coupled POs~\cite{strinati2022hyperspinmachine}. For Ising machines (orange line), whose energy landscape is sketched in Fig.~\ref{fig:hyperrcimpaperfigure2}\textbf{a} as the black wavy line, Ising spins undergo discrete flips, exploring local minima at decreasing energy. During this exploration, local minima trap the system, preventing the reach of the Ising GS. On the contrary, the CHYM converges to an energy below the Ising GS (blue line). Remarkably, since Ising states are not fixed points of the hyperspin dynamics, the hyperspin evolution is driven straight towards the hyperspin GS and cannot be trapped in Ising local minima. When dimensional annealing is performed from the hyperspin GS, the Ising GS is approached from below (red line). The combined effect of being insensitive to Ising local minima and approaching the Ising GS from below is at the core of the enhanced performance of the CHYM with dimensional annealing, as detailed hereafter.

We use the map in Eq.~\eqref{eq:hyperspinmap2} for the numerical simulation of the CHYM and validate the picture in Fig.~\ref{fig:hyperrcimpaperfigure1}\textbf{b} for $N=100$. We focus on $\mathbf{J}_N$ describing fully-connected random sparse graphs with density $20\%$ and, where nonzero, $J_{pq}=\pm J$ with $J>0$ and sign randomly chosen with equal probability. The dimensional annealing in the CHYM is implemented in the FPGA by varying $\mathbf{C}$ in time as $C_{jl}\rightarrow\alpha_{\mu,\tau}C_{jl}$ with $\mu=1+(j-1){\rm mod}(D)$, where the time-dependent functions $\alpha_{\mu,\tau}$ encode the annealing protocol as detailed in the supplementary material.

The CHYM has two sources of nonlinearity: the $\chi^{(2)}$ NLM (local, with nonlinear constant $\kappa L$ being $L$ the propagation length) and the FPGA (non-local, with nonlinear constant $\sim\beta b^2$) coupling all the POs in each hyperspin. Previous studies argued that local nonlinearities are crucial to the functioning of an Ising machine, while non-local ones are detrimental~\cite{PhysRevLett.126.143901,Ben-Ami:23}. In contrast, the hyperspin machine works with only non-local nonlinearities~\cite{strinati2022hyperspinmachine}. Therefore, we set (I) $\kappa L=10^{-3}$ for the CHYM, (II) $\kappa L=10^{-1}$ for the CHYM with dimensional annealing, with $\beta b^2=25$ in both cases in order to have a dominating FPGA nonlinearity, and (III) $\kappa L=10^{-1}$ and $\beta=0$ for the CIM.

In Fig.~\ref{fig:hyperrcimpaperfigure1}\textbf{b}, we see that (I) the CHYM with $D=2$ rapidly converges to the XY GS energy (relative deviation below $0.14\%$), (II) the CHYM with dimensional annealing first converges close to the XY GS and then to the Ising GS from lower energy, and (III) the CHYM used as a CIM ($D=1$) converges to the Ising GS from above~\footnote{Ising and XY GS energies are found by using a Monte-Carlo algorithm and python \texttt{scipy.optmize} \texttt{differential\_evolution} minimizer, respectively.}. These results are in agreement with Ref.~\cite{strinati2022hyperspinmachine}, demonstrating that the CHYM in Fig.~\ref{fig:hyperrcimpaperfigure1} as an implementation of the hyperspin machine.

The remarkable outcome is that existing implementations of CIMs can be adapted into CHYMs by a FPGA software modification, which enables to simulate general continuous spin models and novel annealing strategies using off-the-shelf experimental setups.

\begin{figure}[t]
\centering
\includegraphics[width=8.6cm]{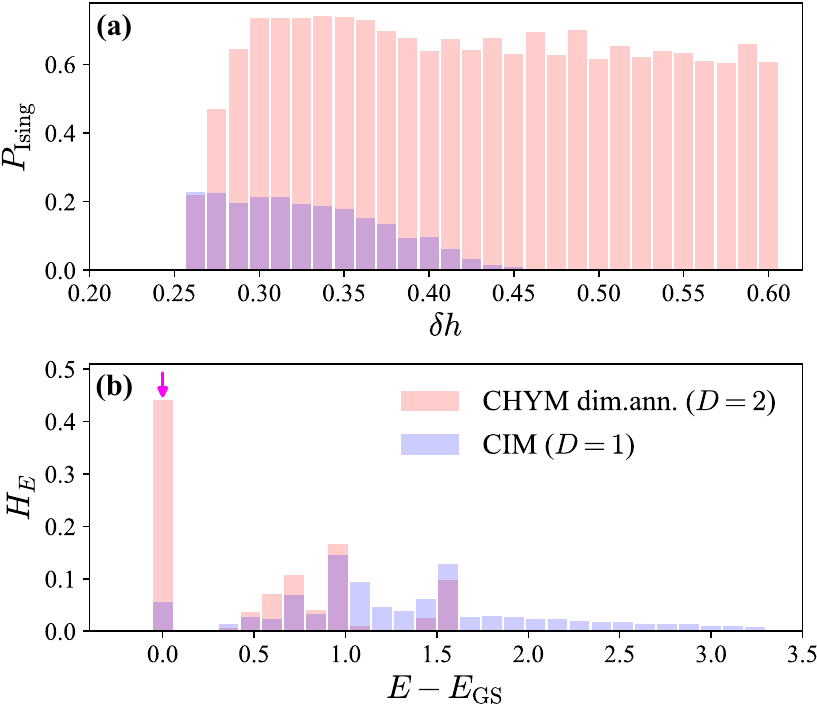}
\caption{Typical distribution of \textbf{(a)} success probability to solve the Ising model $P_{\rm Ising}$ as a function of $\delta h$ and \textbf{(b)} histogram $H_E$ of energy levels as a function of $E-E_{\rm GS}$, comparing the data from the simulation of the CIM (blue) and those from the CHYM with dimensional annealing (red). The magenta arrow marks the GS value $E=E_{\rm GS}$. Compared to the CIM, the CHYM machine with dimensional annealing provides (i) $P_{\rm Ising}$ almost independent of $\delta h$ (where nonzero) and with higher maximum value, and (ii) higher probability $H_E$ to converge to low-energy Ising states. Data are shown for a random graph with $N=100$ for illustration purposes (supplementary material reports comprehensive results).}
\label{fig:hyperrcimpaperfigure3}
\end{figure}

We now move to the statistical and scaling analysis of the performance comparison of the CHYM working as a CIM ($D=1$) and as an XY machine with dimensional annealing ($D=2$). The performance is quantified by running for fixed parameters the CHYM map $N_{\rm r}$ times and computing at the steady state two quantities: (i) The success probability $P_{\rm Ising}$ and (ii) the histogram $H_E$ of energy levels. Here, $P_{\rm Ising}$ counts the fraction of the $N_{\rm r}$ repetitions in which the CHYM converges to the Ising GS energy $E_{\rm GS}$. Instead, $H_{E}$ is the number of times a given Ising energy level $E$ is found.

Previous work reported on the strong dependency of $P_{\rm Ising}$ on both system parameters and coupling matrix~\cite{1805.05217,PhysRevLett.126.143901,strinati2022hyperspinmachine}. To have a reliable statistics for both values of $D$, we simulate the dynamics up to $\tau_{\rm max}=15\times10^4$ for $N_g=100$ sparse graphs, and for $N$ ranging from $10$ to $100$. We repeat the simulations $N_{\rm s}$ values of $h=h_{\rm th}(1+\delta h)$ by scanning the relative deviation $\delta h$ from threshold $h_{\rm th}$. We have $N_{\rm s}=51,21$ for $D=1,2$, respectively and $N_{\rm r}=100$.

\begin{figure}[t]
\centering
\includegraphics[width=8.6cm]{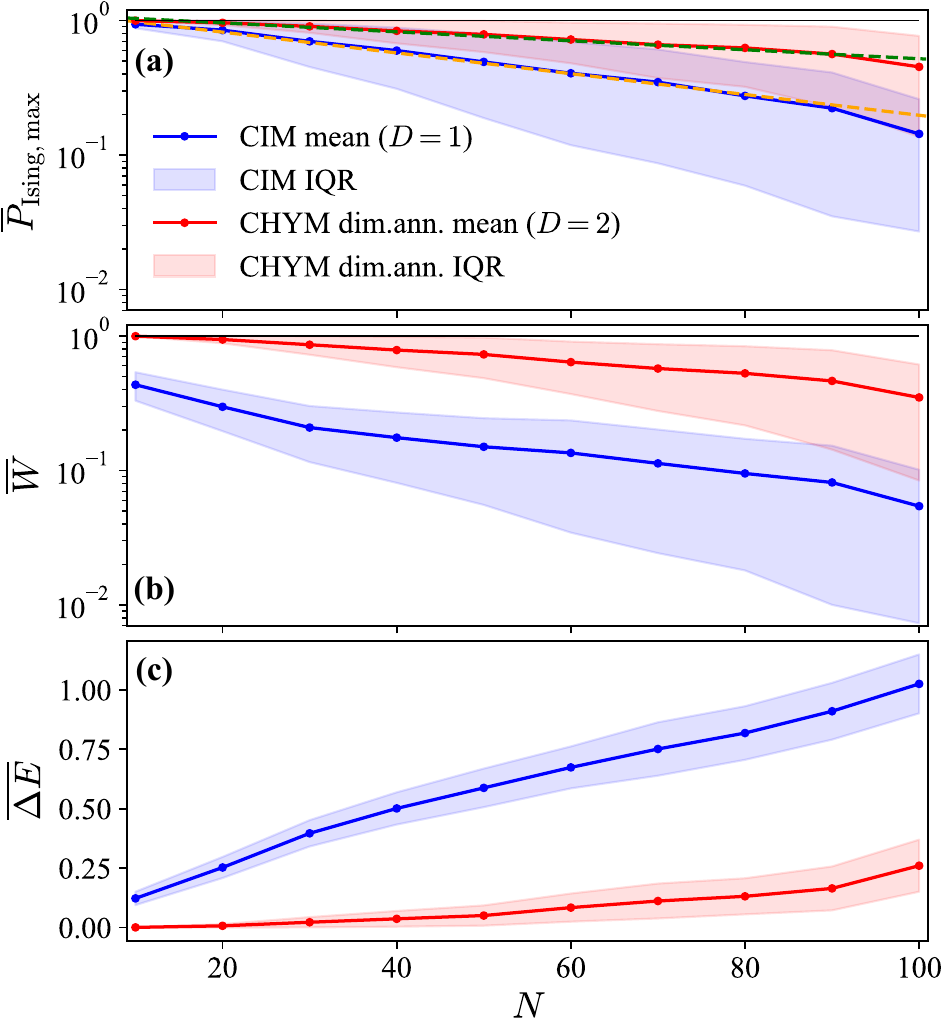}
\caption{Finite-size scaling of \textbf{(a)} $\overline{P}_{\rm Ising,max}$, \textbf{(b)} $\overline{W}$, and \textbf{(c)} $\overline{\Delta E}$, statistically quantifying the advantages of the CHYM with dimensional annealing over the CIM: (i) higher probability to solve the Ising model, (ii) reduced dependency on the pump amplitude, and (iii) enhanced convergence to low-energy Ising states (see also Fig.~\ref{fig:hyperrcimpaperfigure3}). Mean value (points) and IQR (shaded areas) are computed by averaging over $N_g=100$ graphs for each value of $N$ and $D$ (see text). Dashed lines in top panel are exponential functions $f_D(N)=b_D\,e^{-a_DN}$ with $a_1\simeq18\times10^{-3}$ (orange) and $a_2\simeq8\times10^{-3}$ (green). Notice the log-linear scale in panels \textbf{(a)} and \textbf{(b)}}
\label{fig:hyperrcimpaperfigure4}
\end{figure}

Figure~\ref{fig:hyperrcimpaperfigure3} exemplifies the behaviour of $P_{\rm Ising}(h)$ and $H_E$. By comparing $D=1$ (blue) and $D=2$ (red), we observe three striking differences: (i) $P_{\rm Ising}$ for $D=2$ has higher maximum value over the pump scans; (ii) $P_{\rm Ising}$ for $D=2$ has a much weaker dependency on $h$; (iii) $H_E$ is considerably closer to $E_{\rm GS}$. The dimensional annealing running in the FPGA boosts the performance of the CHYM in terms of enhanced accuracy in the Ising GS and reduced sensitivity to the system hyperparameters.

The analysis suggests three figures of merit extracted from the data in Fig.~\ref{fig:hyperrcimpaperfigure3}. If $\Delta h$ is the range of $h$ where $P_{\rm Ising}(h)$ is nonzero, we define:
\begin{itemize}
\item[(i)] $P_{\rm Ising,max}={\rm max}_h\{P_{\rm Ising}(h)\}\leq1$, giving the maximum value of $P_{\rm Ising}(h)$;
\item[(ii)] $W=S^{-1}\sum_{h\in\Delta h}P_{\rm Ising}(h)\leq1$, quantifying $P_{\rm Ising}(h)$ and its sensitivity on $h$, being $S=\sum_{h\in\Delta h}$ a normalization factor;
\item[(iii)] $\Delta E=\sum_E(E-E_{\rm GS})H_E$, i.e., the average energy deviation from the GS. When $\Delta E=0$ the CHYM finds the Ising GS with probability $1$. Increasing values of $\Delta E$ signal a reduced convergence to low-energy states.
\end{itemize}
Figure~\ref{fig:hyperrcimpaperfigure4} shows the finite-size scaling of $\overline{P}_{\rm Ising,max}$, $\overline{W}$, and $\overline{\Delta E}$, defined respectively as the mean values of $P_{\rm Ising,max}$, $W$, and $\Delta E$, over the $N_g$ graphs. Data are shown as connected points (blue and red for $D=1,2$, respectively), and the statistical uncertainty is the interquartile range (IQR, shaded areas).

The decay $\overline{P}_{\rm Ising,max}$ in Fig.~\ref{fig:hyperrcimpaperfigure4}\textbf{a} is well captured by an exponential $\sim e^{-a_DN}$ in both cases (orange and green dashed lines), with $a_1/a_2\simeq2.32$. For $D=1$, this exponential trend is in agreement with Ref.~\cite{1805.05217}. The implication of this result is twofold: First, the decay of $\overline{P}_{\rm Ising,max}$ for the CHYM with dimensional annealing is also exponential, and second, its decay rate is halved with respect to the CIM. In other words, the CHYM with dimensional annealing shows a decay of $\overline{P}_{\rm Ising,max}$ that is exponentially slower (i.e., the performance is exponentially improved) by a factor $e^{(a_1-a_2)N}$.

The scaling of $\overline{P}_{\rm Ising,max}$ provides only informations on the optimal success probability. It is complemented by the trend of $\overline{W}$ in Fig.~\ref{fig:hyperrcimpaperfigure4}\textbf{b}, which gives a comprehensive information on the dependence on $h$ of the success rate. We see that $\overline{W}$ for $D=2$ is almost one order of magnitude larger than for $D=1$, i.e., on average, the dimensional annealing makes $P_{\rm Ising}(h)$ significantly less sensitive to the specific value of $h$. This fact has a remarkable advantage, both in simulations and experiments: The CHYM with dimensional annealing does not require a fine calibration of the pump to operate in the optimal regime. We ascribed the large IQR to the strong graph dependency of the considered figures of merit.

Figure~\ref{fig:hyperrcimpaperfigure4}\textbf{c} shows $\overline{\Delta E}$. As evident, $\overline{\Delta E}$ for $D=1$ starts from $\sim0.1$ at $N=10$, and rapidly increases with $N$ reaching $\sim1.0$ for $N=100$. In contrast, for $D=2$, $\overline{\Delta E}$ is close to zero up to $N=60$, showing a slower increase compared to $D=1$, reaching $\sim0.1$ at $N=100$. The reported data provide clear evidence that the CHYM increases the tendency to converge to low-energy Ising states in comparison with the CIM.

In conclusion, an extensive statistical analysis shows that the hyperspin heuristics provide exponential improvement in the finite-size scaling of Ising machines. Also, a simple software modification to the feedback mechanism in the CIM enables an experimental implementation through an effective nonlinear coupling between POs. For a family of random graphs, we unveiled that the resulting CHYM can handle general high-dimensional models and dimensional annealing.

We identified three figures of merit as benchmarks: (i) Maximum success rate, (ii) Sensitivity of the success rate to pump amplitude value, and (iii) Convergence accuracy to the Ising energy ground state. High-performance-computing simulations furnish the scaling of the figures of merit averaged over hundreds of graphs. Compared with the CIM, the CHYM has a more reliable and successful convergence to the Ising ground state and does not require a fine calibration of the pump amplitude. Our proposal turns state-of-the-art CIMs into hyperspin machines boosting their performance. The dimensional annealing exponentially improves the figures of merit in the accuracy of the optimal solution and also in the sensitivity to hyperparameters.

The application of hyperspins and dimensional annealing is not limited to CIMs but extends to other hardware platforms, including quantum devices. The crucial recipe is engineering the non-local couplings and the nonlinear loss. The high-dimensional embedding not only may accelerate by orders of magnitude classical and quantum computing, but also trigger the development of new algorithms for large-scale optimization and machine learning.

\begin{acknowledgements}
 We acknowledge the CINECA award under the ISCRA initiative for the availability of high-performance-computing resources and support. Numerical simulations of the CHYM are performed using a dedicated C-language code exploiting multiprocessing on the CINECA GALILEO100 supercomputer over $400$ CPUs. C.C. acknowledges financial support from CN1 Quantum PNRR MUR CN\_0000013 HPC.
\end{acknowledgements}

%\bibliography{bibliography_parametric_oscillators}


\end{document}